\documentclass[times, twoside, watermark]{arxiv}
\usepackage{blindtext}
\usepackage{booktabs}
\usepackage{array}
\usepackage{tabularx}
\newcolumntype{Y}{>{\raggedright\arraybackslash}X}

\leadauthor{Dominik Jasi\'owka} 

\begin{document}

\title{A Real-Time Low-Cost Platform for Non-Contact Respiratory Rate Monitoring: A Comparative Evaluation of RGB Optical Flow and Time-of-Flight Depth Sensing}
\shorttitle{Low-Cost Platform for Non-Contact Respiratory Rate Monitoring}


\author[1]{Dominik Jasi\'owka}

\author[2,3]{Patryk Ci\k{e}\.zarek}

\author[2,3]{Joanna Zyznawska}

\author[1,\Letter]{Ludwin Molina Arias}

\affil[1]{Department of Biocybernetics and Biomedical Engineering, Faculty of Electrical Engineering, Automatics, Computer Science and Biomedical Engineering, AGH University of Krakow, 30-059 Krakow, Poland}

\affil[2]{Institute of Physiotherapy, Faculty of Health Science, Jagiellonian University Medical College, 31-126 Krakow, Poland}

\affil[3]{Motor Function Research Laboratory, Center for the Development of Therapies for Civilization and Age-Related Diseases, Jagiellonian University Medical College, 31-126 Krakow, Poland}

\maketitle


\begin{abstract}
Respiratory rate (RR) is one of the earliest indicators of clinical deterioration, which makes its continuous measurement clinically valuable. In routine practice, however, RR is monitored almost exclusively with contact-based sensors, which require skin attachment, may alter natural breathing, and are poorly tolerated by vulnerable groups such as newborns and elderly patients. Non-contact sensing removes these constraints, with RGB video and depth imaging among its most accessible modalities. In this work, we present a low-cost, real-time platform based on a single multimodal sensor that provides both RGB and time-of-flight (ToF) data streams, implemented on a Raspberry Pi 5 edge device. The system integrates three RGB optical-flow pipelines, namely Farneb\"ack, Horn--Schunck, and Lucas--Kanade, together with a depth-based pipeline incorporating region-of-interest stabilization and distance-derived signal extraction, enabling the two modalities to be evaluated independently under the same experimental protocol. The proposed approach was evaluated on twelve healthy volunteers following a controlled protocol, including normal breathing, hyperventilation, and breath-holding, using a piezoresistive respiratory belt as a reference. RGB-based methods achieved lower RR estimation errors, with Lucas--Kanade attaining the lowest median MAE ($3.83$~BPM), whereas the ToF-based pipeline, although less accurate for RR estimation (median MAE $6.24$~BPM), reached a higher apnea detection sensitivity ($0.66$ vs $0.51-0.55$). These findings highlight the complementary strengths of both modalities and support their integration into unified low-cost platforms.
\end {abstract}

\begin{keywords}
apnea detection, computer vision, depth sensing, edge computing, respiratory monitoring
\end{keywords}

\begin{corrauthor}
arias\at agh.edu.pl
\end{corrauthor}

\section{Introduction}
\label{sec:introduction}
Respiratory rate (RR) is one of the most sensitive indicators of physiological deterioration in clinical practice. Changes in breathing often precede alterations in heart rate or blood pressure by several hours, allowing for the early detection of conditions such as cardiopulmonary failure or sepsis \cite{Addison2021,Cretikos2008}. Consequently, RR is a key component of early-warning systems (e.g. National Early Warning Score), where deviations from normal adult resting values of 12--20~breaths per minute (BPM) indicate elevated risk \cite{romano2021,Smith2019}. Importantly, normal ranges vary with age: newborns can exceed $40$~BPM, decreasing within the first years of life \cite{fleming2011}, while older adults ($>$65 years) can reach $28$~BPM at rest \cite{RodrguezMolinero2013}. These differences highlight the need for monitoring systems that remain reliable across populations and capture both slow and rapid breathing dynamics.

Despite its clinical importance, RR is still predominantly measured using contact-based techniques. Common solutions include respiratory belts, strain sensors, inertial measurement units (IMUs), surface electromyography (sEMG), and indirect estimation from photoplethysmography (PPG) or electrocardiography (ECG) signals \cite{Hussain2023,Cesareo2018,charlton2018,Kumar2022}. Although these methods can provide accurate measurements under controlled conditions, they share several practical limitations. The need for body contact can cause discomfort, particularly in vulnerable populations such as newborns or elderly patients, and can alter natural breathing patterns due to sensor awareness. In addition, these systems often require wiring, consumables, and trained personnel, which restricts their applicability for long-term monitoring, home care, and large-scale deployment while also increasing the risk of contamination. These limitations have motivated the development of non-contact alternatives that can preserve comfort while maintaining clinical reliability.

Non-contact RR monitoring has been explored using three main sensing modalities: RGB vision, thermal imaging, and active sensing (depth or radar). Vision-based methods extract breathing signals from subtle chest movements or pixel intensity variations and may include remote photoplethysmography (rPPG) techniques \cite{verkruysse2008}. They benefit from low cost and high resolution but remain sensitive to illumination and occlusions \cite{Massaroni2018,romano2021}. Thermal imaging, by contrast, detects cyclic temperature changes near the nostrils, allowing operation in darkness, though at a higher cost \cite{Massaroni2021}. Active sensing techniques, such as depth cameras and radar, directly measure chest displacement and are robust to lighting and skin conditions, albeit at moderate resolution.

Although these approaches have demonstrated promising results, two key gaps remain. First, most studies evaluate individual sensing modalities in isolation, making it difficult to establish fair comparisons under consistent experimental conditions and reference standards. Second, many proposed solutions rely on expensive hardware setups and workstation-level processing, limiting their translation to real-world and resource-constrained environments. In this context, low-cost embedded systems that integrate multiple sensing modalities while maintaining reliable performance remain largely underexplored.

To address these challenges, this work proposes a unified, low-cost framework for non-contact respiratory monitoring based on a compact multimodal sensor integrating RGB and depth data on a Raspberry Pi~5. The system evaluates three RGB-based pipelines based on optical flow alongside a depth-based approach under the same cohort and experimental protocol, using the same respiratory-belt reference system, while processing each modality independently to enable a controlled and unbiased comparison. This unified design facilitates a direct assessment of the modalities in terms of RR estimation accuracy, waveform agreement, phase-specific behavior, apnea detection capability, and computational performance in an embedded setting.

\section{Related Work}
\label{sec:related-work}

Vision-based approaches have emerged as a promising alternative for non-contact respiratory monitoring, aiming to recover the breathing signal either from subtle cyclic variations in pixel intensity or from the motion of the chest wall. Early studies demonstrated that, under controlled conditions, these methods can achieve high accuracy. For instance, Massaroni et al.~\cite{Massaroni2018} reported a near-zero bias of $0.01 \pm 1.81$~BPM using an intensity-based approach in six volunteers. Similarly, Romano et al.~\cite{romano2021} validated an optical-flow pipeline in 24 subjects, obtaining a bias of $-0.03 \pm 1.38$~BPM at a distance of two meters. However, these results also revealed an intrinsic limitation of RGB-based methods, namely their sensitivity to illumination variations, motion artifacts, and partial occlusion of the thoracic region.

To overcome these constraints, subsequent research shifted towards identifying alternative regions of interest that remain informative even when the chest is not fully visible. This led to the exploration of micro-motions of the head mechanically coupled with respiration \cite{schrumpf2019}, the suprasternal notch, and lateral body profiles \cite{Mateu-Mateus2020}.

Moving beyond controlled laboratory settings, Zhu et al.~\cite{zhu2019} demonstrated that vision-based approaches can retain acceptable performance under more realistic conditions, reporting a root-mean-square (RMS) error of $2.10 \pm 1.64$~BPM against polysomnography in 50 adults during overnight monitoring, with limited sensitivity to sleep position, sleep stage, or bedding coverage. More recent research has further advanced the field by replacing hand-crafted pipelines with learning-based solutions, such as motion magnification combined with convolutional neural networks \cite{brieva2020} and hybrid architectures integrating optical-flow and monocular-depth cues \cite{Ghezzi2025}.

In parallel, thermal imaging has been investigated as a complementary modality, utilizing cyclic temperature variations around the nostrils to estimate the respiratory signal. Because it does not depend on external illumination, thermal imaging enables operation in complete darkness \cite{Massaroni2021}. Clinical studies have reported competitive accuracy, with Alves et al.~\cite{alves2025} achieving a mean absolute error (MAE) between $0.64$ and $0.91$~BPM in overnight hospital recordings. Methodological developments in this area have focused on robust nasal-region tracking \cite{Chiang2016,nagata2018} and, more recently, on deep spatio-temporal models \cite{mozafari2024}. Nevertheless, the high cost of long-wave infrared sensors compared to consumer-grade RGB cameras continues to limit their widespread adoption, particularly in ambulatory or home-care settings.

Active sensing technologies provide an alternative route to overcoming the limitations of passive vision systems. Among them, time-of-flight (ToF) depth cameras are particularly relevant, as they directly measure thoracic displacement by capturing distance maps that are largely independent of ambient lighting and skin pigmentation. This capability enables the extraction of stable respiratory waveforms and the detection of events such as apnea \cite{Addison2021}. LiDAR-based implementations have further extended this principle \cite{hill2021}, incorporating thermally guided region selection \cite{binnesar2022} and demonstrating robustness from multiple viewpoints, including frontal, dorsal, and lateral configurations \cite{rinchi2024}. Radar-based systems constitute another complementary class of illumination-independent approaches that are capable of operating through clothing or light bedding. These include a Frequency-Modulated Continuous-Wave (FMCW) radar for simultaneous monitoring of respiration and heart rate \cite{lv2021}, a validated Ultra-wideband impulse radio range (UWB-IR) radar in sensitive populations such as newborns \cite{kim2019}, and monopulse radar systems combined with source separation techniques for multi-subject monitoring \cite{islam2020}.

Despite substantial methodological advances across RGB, thermal, and active sensing approaches, the current body of literature remains limited in aspects that affect the translation of these solutions into practical systems. Existing studies predominantly focus on a single sensing modality evaluated in isolation; therefore, the reported performance is strongly influenced by dataset-specific factors, acquisition setups, and evaluation protocols. This makes it difficult to establish fair comparisons between modalities, while direct side-by-side evaluations of RGB and depth-based methods under identical conditions---same subjects, protocol and reference signal---remain scarce. At the same time, many proposed systems rely on dedicated hardware configurations and workstation-level processing, which restrict their applicability in real-world scenarios that require portability, low cost, and real-time performance. Although individual components of the pipeline, such as optical flow, have been extensively studied, their integration within a unified, resource-constrained embedded platform has received comparatively limited attention.

In this context, the present work introduces a unified multimodal framework that jointly addresses these limitations. Unlike previous studies, the proposed system integrates RGB and ToF depth sensing within a single compact sensor, enabling a direct comparison of both modalities under identical acquisition conditions. These modalities were selected because they provide complementary information while remaining suitable for low-cost embedded deployment. Optical flow estimates thoracic motion directly from RGB imagery without requiring task-specific training data \cite{romano2021,Mateu-Mateus2020}, whereas ToF sensing provides direct measurements of thoracic displacement and is largely insensitive to illumination and skin pigmentation \cite{Addison2021,hill2021}. Moreover, acquiring both streams from the same device ensures a common viewpoint and hardware setup, which are essential for a controlled comparison.

To assess the impact of motion estimation strategies within the RGB modality, the evaluation includes three representative optical-flow approaches: Lucas--Kanade, Horn--Schunck, and Farneb\"ack. These methods span the principal formulations of optical flow, namely sparse feature-based, dense variational, and dense polynomial-expansion approaches, respectively, enabling a systematic analysis of algorithmic variability within a unified framework. While the present study focuses on these optical-flow methods and ToF depth sensing, the proposed architecture remains modular and can accommodate alternative motion estimation algorithms or depth sensing technologies.

Beyond the comparison of sensing modalities and motion estimation approaches, the platform is designed for real-time operation on a Raspberry Pi~5, demonstrating that multimodal respiratory monitoring can be achieved without specialized hardware or high-performance computing resources. Furthermore, the evaluation extends beyond conventional RR estimation accuracy by incorporating waveform-level agreement, phase-specific analysis, apnea detection performance, and computational efficiency, providing a comprehensive characterization of respiratory monitoring performance.

\section{Materials and Methods}
\label{sec:materials-methods}

\subsection{System Architecture and Data Acquisition}
\label{sec:mm-hardware}

The proposed platform was developed around a single low-cost multimodal sensor (MaixSense A075V), which integrates an RGB camera and a ToF depth camera and is connected to a Raspberry Pi 5 responsible for data acquisition, processing, and visualization via a lightweight server. Data were retrieved through the sensor's HTTP interface over a local network. In the RGB configuration, JPEG-compressed color images were decoded at approximately 30~fps. In the depth configuration, the ToF camera operated in raw 16-bit mode, providing uncompressed distance measurements in millimeters. A longer exposure time was selected to reduce measurement noise, resulting in an operating frame rate of approximately 15~fps at a resolution of $320\times240$ pixels.

\subsection{Respiratory Monitoring Pipelines}

\begin{figure*}[!t]
\centering
\includegraphics[width=0.8\linewidth]{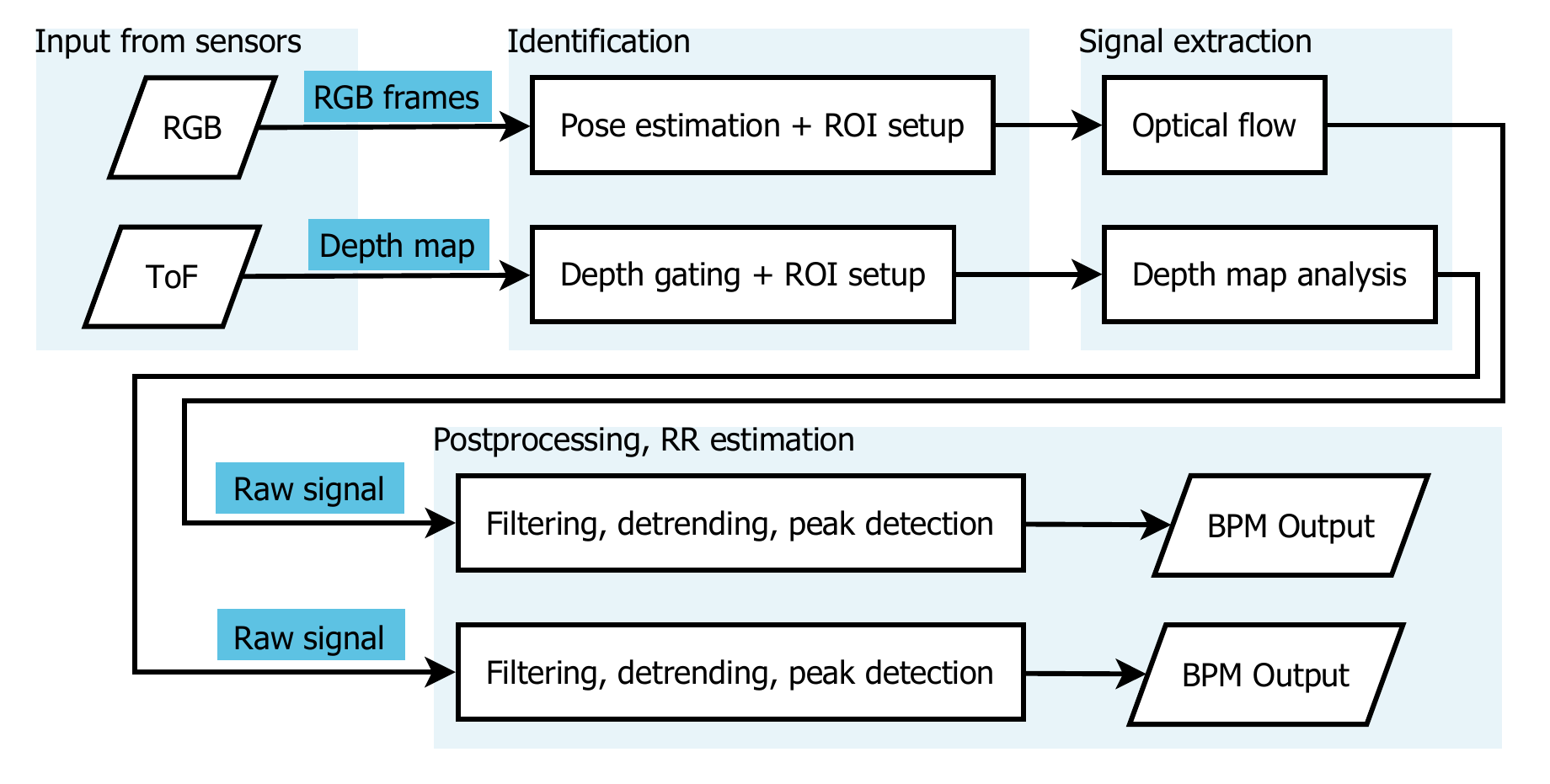}
\caption{Overview of the proposed low-cost platform and the two independent respiratory monitoring pipelines, each producing its own RR estimate.}
\label{fig:dataflow}
\end{figure*}

Both sensing modalities follow the same workflow: subject identification, respiratory signal extraction, and respiratory analysis. Each modality is processed independently, producing its own respiratory waveform and RR estimate. The complete architecture is shown in Fig.~\ref{fig:dataflow}.

\subsubsection{RGB Pipeline}

\begin{figure*}[!t]
\centering
\includegraphics[width=0.8\linewidth]{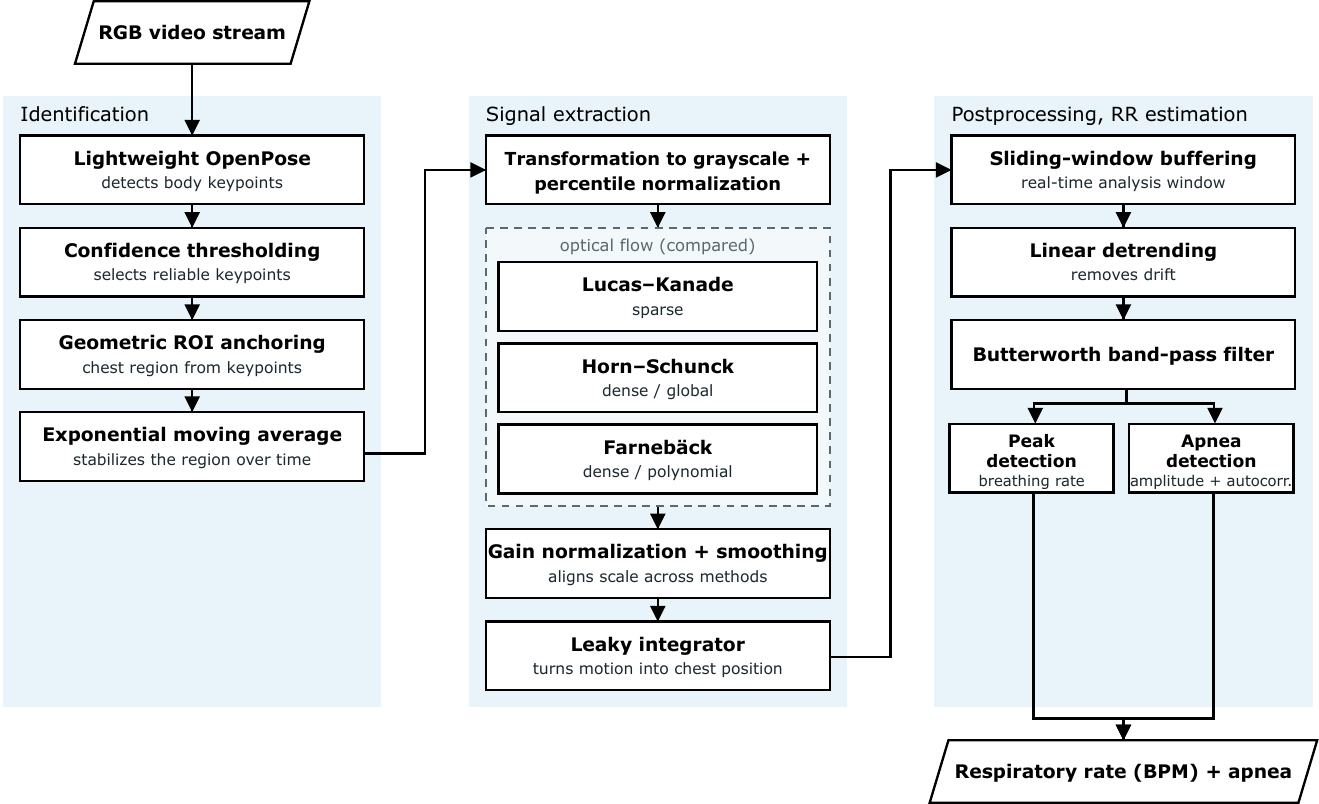}
\caption{Overview of the RGB pipeline. The three optical-flow methods are applied in parallel to the same thoracic ROI, each yielding an independent respiratory signal.}
\label{fig:rgb-pipeline}
\end{figure*}

Figure~\ref{fig:rgb-pipeline} summarizes the RGB processing chain.

\paragraph{Subject Identification and Region-of-Interest (ROI) Extraction}

The thoracic region is localized using a pretrained Lightweight OpenPose network \cite{osokin2019}, derived from the original OpenPose framework \cite{cao2021}. Each frame is resized to $192\times192$ pixels, and the keypoint locations are estimated from the corresponding confidence heatmaps, retaining only detections above a predefined confidence threshold. The thoracic ROI is constructed from geometric relationships between the neck and shoulder landmarks, scaled according to shoulder width and upper-body proportions using empirically tuned factors, while physiologically implausible aspect ratios are rejected. To reduce computational cost and temporal jitter, detections are periodically updated and smoothed using an exponential moving average.

\paragraph{Respiratory Signal Extraction}

Respiratory motion is estimated from the transverse displacement of the thoracic ROI between consecutive frames. Following grayscale conversion and percentile-based contrast normalization, motion is quantified using three optical-flow methods: Lucas--Kanade, Horn--Schunck, and Farneb\"ack. All methods are applied independently to the same thoracic ROI and the same pair of consecutive frames, yielding separate respiratory waveforms and RR estimates. Their outputs are not combined, allowing each method to act as an independent estimator whose performance is directly compared in Section~\ref{sec:results}. All three approaches rely on the brightness-constancy assumption, which states that the intensity of a moving surface point remains unchanged between consecutive frames. Under this assumption, the image brightness function satisfies the following equation:
\begin{equation}
I_x\,u + I_y\,v + I_t = 0
\label{eq:ofc}
\end{equation}
where $(u,v)$ denotes the horizontal and vertical components of the optical flow vector, and $I_x$, $I_y$, and $I_t$ represent the partial derivatives of the image intensity with respect to $x$, $y$, and $t$, respectively.

\emph{Lucas--Kanade} method \cite{Lucas1981} assumes locally constant motion within a small spatial neighborhood around each tracked point and estimates sparse motion vectors by solving a local least-squares problem on selected corner features \cite{JianboShi1994}. The respiratory signal is derived as the median vertical displacement of the tracked features.

\emph{Horn--Schunck} method \cite{Horn1981} computes a dense flow field by minimizing a global energy functional that balances brightness constancy and spatial smoothness:
\begin{equation}
E = \iint \left[ (I_x u + I_y v + I_t)^2 + \alpha^2 \left( \lVert \nabla u \rVert^2 + \lVert \nabla v \rVert^2 \right) \right] dx\,dy
\label{eq:hsenergy}
\end{equation}
where $\alpha$ denotes a regularization parameter, and $\nabla u$ and $\nabla v$ represent the spatial gradients of the optical flow components.

The flow is estimated iteratively using Jacobi relaxation, and the respiratory signal corresponds to the spatial mean of the transverse component.

\emph{Farneb\"ack} method \cite{Farnebck2003} models local image neighborhoods through quadratic polynomial expansions and estimates dense motion from changes in polynomial parameters across frames. As in Horn--Schunck, the respiratory signal is computed from the mean transverse flow component.

Because all optical-flow outputs represent velocity-like quantities, they are normalized, smoothed, and converted into a displacement-like respiratory waveform through a leaky integrator:
\begin{equation}
s(t) = \gamma\,s(t-1) + \tilde{v}(t)\,\Delta t
\label{eq:leaky}
\end{equation}
where $\tilde{v}(t)$ denotes the smoothed velocity estimate, $\gamma = 0.98$ is the decay factor, $\Delta t$ is the interval between frames, and $s(t)$ is the one-dimensional respiratory signal.

\paragraph{Respiratory Signal Processing}

Respiratory analysis is performed using a sliding-window approach. Slow baseline variations are first removed via linear detrending, followed by zero-phase forward--backward filtering with a fourth-order Butterworth band-pass filter (0.08--2.5 Hz).

Apnea detection relies on two complementary characteristics of the filtered signal: amplitude, quantified as the standard deviation within a short sliding window, and periodicity, assessed through the normalized autocorrelation function. A segment is classified as breathing only when both measures exceed their respective thresholds, and individual decisions are temporally withdrawn to improve robustness against transient fluctuations.

Whenever a valid breathing pattern is detected, the RR is estimated from the intervals between successive signal peaks identified under minimum prominence and spacing constraints. In parallel, evidence of respiratory absence is accumulated over time using a leaky integrator with hysteresis:
\begin{equation}
I_a(t) =
\begin{cases}
I_a(t-1) + c_a, & \text{apnea} \\[3pt]
\gamma_a\,I_a(t-1), & \text{breathing}
\end{cases}
\label{eq:apnea}
\end{equation}
where $I_a(t)$ is the accumulation state of apnea in frame $t$, $c_a$ is the fixed increment added for each frame classified as apnea, and $\gamma_a < 1$ is the decay factor applied during frames classified as breathing.

\subsubsection{ToF Pipeline}
\label{sec:mm-tof}

\begin{figure*}[!t]
\centering
\includegraphics[width=0.8\linewidth]{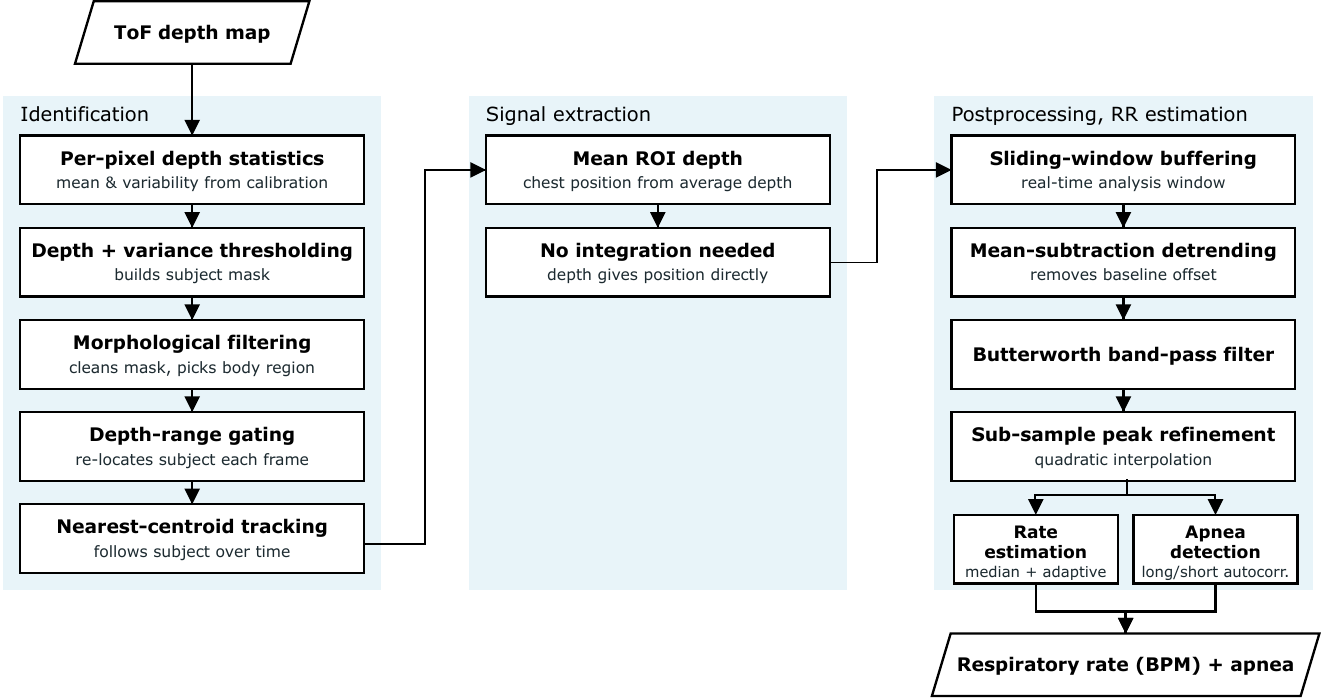}
\caption{Overview of the ToF pipeline.}
\label{fig:tof-pipeline}
\end{figure*}

The ToF processing chain is shown in Fig.~\ref{fig:tof-pipeline}.

\paragraph{Subject Identification and ROI Extraction}

Unlike the RGB approach, the ToF pipeline identifies the subject directly in the depth image. During calibration, pixel-wise temporal mean and standard deviation maps are computed from a set of depth frames; the mean depth separates the subject from the background, while the standard deviation highlights respiratory motion. A binary mask obtained by depth and variability thresholding is refined through morphological operations. The largest connected component is then retained, and a fixed-size ROI is initialized around its centroid. During monitoring, the subject is tracked by selecting depth pixels near the reference distance and identifying the connected component closest to the previous location, with the ROI position and reference depth updated using exponential moving averages.

\paragraph{Respiratory Signal Extraction}

For each frame, the respiratory signal is computed as the average depth within the ROI. Because ToF measurements provide an absolute distance in millimeters, the resulting signal directly reflects the thoracic displacement and does not require integration.

\paragraph{Respiratory Signal Processing}

The depth signal is processed using the same respiratory-analysis framework as the RGB pipeline, the main differences being mean-based detrending and a narrower band-pass filter (0.05--1.5 Hz), reflecting the lower sampling rate and smoother characteristics of the depth signal. To compensate for the reduced temporal resolution, detected peaks are refined through quadratic interpolation of the samples adjacent to each peak, and the RR is computed from recent inter-peak intervals after median-based outlier rejection and adaptive exponential smoothing. Apnea detection additionally compares the periodicity of long-term and short-term windows. When a recent change in the breathing pattern is detected, older samples are discarded to accelerate adaptation and avoid false apnea detections.

\subsection{Study Design and Experimental Protocol}
\label{sec:mm-protocol}

\subsubsection{Participants}
The platform was evaluated in a cohort of twelve healthy adult volunteers without known respiratory disorders or clinically diagnosed lung conditions. All procedures involving human participants were reviewed and approved by the Research Ethics Committee of the Jagiellonian University Medical College (Resolution No.~47/2026, 22~January 2026) and were conducted according to the Declaration of Helsinki. Written informed consent was obtained from all participants prior to inclusion in the study.

The cohort comprised 7 male and 5 female participants, whose demographic characteristics are summarized in Table~\ref{tab:cohort}.
\begin{table}[!t]
\caption{Demographic Characteristics of the Study Cohort ($n = 12$; 7 Male, 5 \mbox{Female})}
\label{tab:cohort}
\centering
\footnotesize
\setlength{\tabcolsep}{3pt}
\begin{tabularx}{\linewidth}{@{}Ycc@{}}
\toprule
\textbf{Characteristic} & \textbf{Mean $\pm$ SD} & \textbf{Range} \\
\midrule
Age [years]                               & $29.0 \pm 11.5$ & 21--54 \\
Body height [cm]                          & $174.7 \pm 9.5$ & 156--187 \\
Body mass [kg]                            & $72.7 \pm 9.6$  & 55--85 \\
Body-mass index [kg/m\textsuperscript{2}] & $23.8 \pm 2.0$  & 21.4--27.8 \\
\bottomrule
\end{tabularx}
\end{table}

\subsubsection{Experimental Protocol}

All recordings were conducted indoors under ambient daylight conditions, with participants seated facing the sensing device and the MaixSense A075V positioned approximately 60 cm in front of the thorax. Each participant completed two separate recording sessions, one using the RGB modality and one using the ToF modality, following the same experimental protocol. Each session had its own corresponding respiratory-belt recording, which served as the reference signal for that modality.

\begin{figure*}[!t]
\centering
{\small\textbf{(a)}}\\[2pt]
\includegraphics[width=0.8\linewidth]{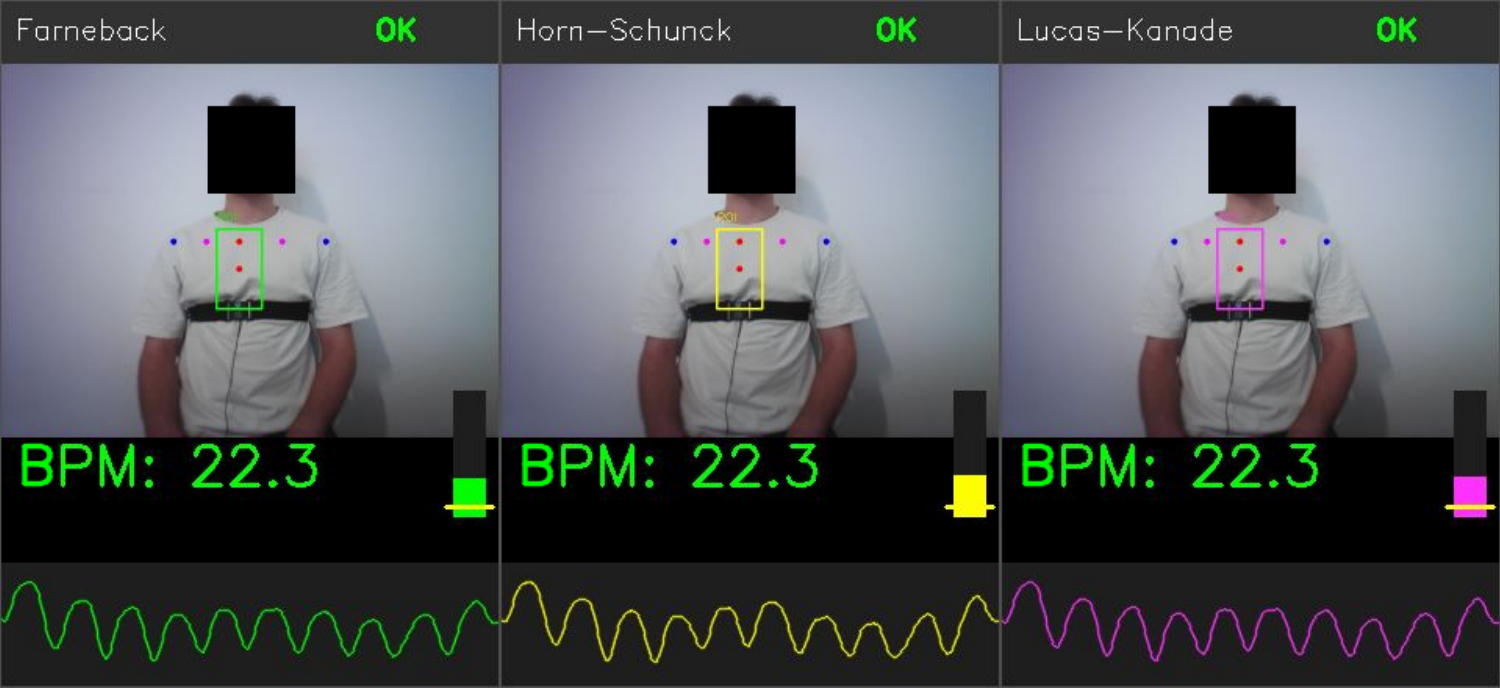}\\[6pt]
{\small\textbf{(b)}}\\[2pt]
\includegraphics[width=0.25\linewidth]{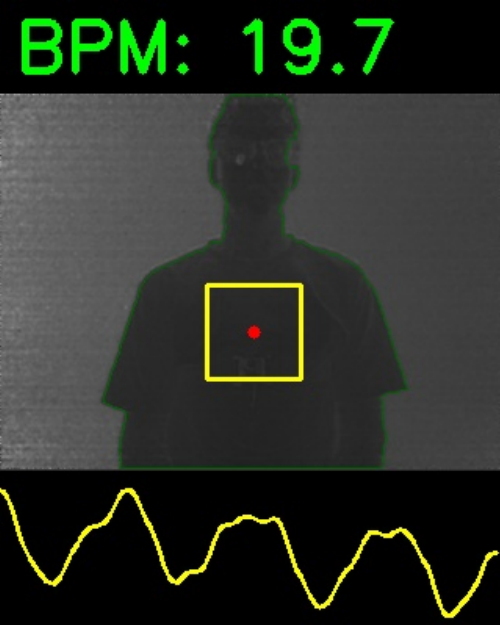}\\[2pt]
\caption{Real-time visualization of the proposed respiratory monitoring system. \textbf{(a)} RGB pipeline showing pose keypoints, thoracic ROI, respiratory waveform, and RR estimated using the three optical flow methods. \textbf{(b)} ToF pipeline showing the tracked ROI, centroid, and depth-derived respiratory waveform with RR estimate.}
\label{fig:views}
\end{figure*}

Each session consisted of a calibration stage during which participants maintained normal breathing while modality-specific parameters were initialized (a statistical depth model of the scene for the ToF pipeline, a baseline respiratory amplitude for the detection of apnea in the RGB pipeline), followed by a structured breathing protocol comprising three consecutive phases: (i) normal resting breathing, (ii) voluntary hyperventilation to induce an elevated RR, and (iii) breath-holding to simulate apnea. Both pipelines provided visual feedback in real-time on the processing stages. Representative outputs, including detected ROI, respiratory waveform, and estimated RR, are shown in Fig.~\ref{fig:views}.

\subsubsection{Data Preparation}

The study includes 24 recordings, one per participant and modality, all processed offline. The RGB and ToF data are treated as independent measurements rather than frame-wise paired signals. To handle the non-stationary structure of the protocol, the breath-hold and hyperventilation segments were manually annotated in each recording, allowing the reference RR to be set to zero during apnea and to be estimated using phase-appropriate parameters during rapid breathing. Since camera-derived and reference signals were acquired with independent devices, both were resampled on a common 20~Hz grid using linear interpolation, normalized to zero mean and unit variance, and temporally aligned by estimating the delay that maximized their cross-correlation.

\section{Results}
\label{sec:results}

\subsection{Real-Time Performance on the Edge Device}
\label{sec:res-perf}

To verify that the proposed pipelines operate within the constraints of a low-cost edge platform, the instantaneous frame rate and CPU load were logged at 1~Hz during a continuous one-minute run on the Raspberry Pi 5, with each modality measured independently. In the RGB configuration, the three optical-flow estimators were executed concurrently within a single process, so the reported throughput corresponds to their combined computational cost. The CPU load is reported as a system-wide average across all four cores, where 100\% corresponds to full utilization. The measured throughput and computational load are summarized in Table~\ref{tab:perf}.
\begin{table}[!t]
\caption{Real-Time Performance of the RGB and ToF Pipelines on the Raspberry Pi 5}
\label{tab:perf}
\centering
\footnotesize
\setlength{\tabcolsep}{3pt}
\begin{tabularx}{\linewidth}{@{}Ycc@{}}
\toprule
\textbf{Metric} & \textbf{RGB} & \textbf{ToF} \\
\midrule
Mean frame rate [fps] & 15.6 & 13.1 \\
Target frame rate [fps] & 30 & 15 \\
Mean CPU load [\%, all cores] & 25.3 & 4.5 \\
Logging duration [s] & 68 & 76 \\
\bottomrule
\end{tabularx}
\end{table}

The RGB pipeline operated below the nominal sensor capability of 30~fps, mainly due to the computational cost of pose estimation and optical-flow processing on a CPU-only platform, although pose inference is performed only once every five frames. The ToF pipeline operated close to its target rate with a much lower computational demand, consistent with the absence of neural-network inference. Therefore, both pipelines meet the timing requirements for continuous respiratory monitoring on the target edge device, with the ToF pipeline retaining significant computational headroom for additional concurrent tasks.

\subsection{Waveform-Level Agreement Evaluation}
\label{sec:res-corr}

\begin{figure}[!t]
\centering
\includegraphics[width=\linewidth]{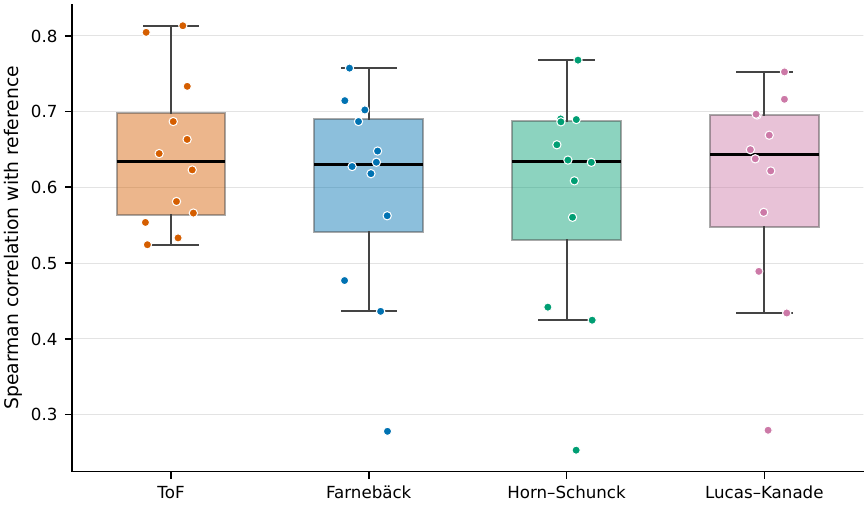}
\caption{Per-subject Spearman correlation between the estimated and the reference (belt) respiratory waveforms (z-score normalized, 20~Hz), by method.}
\label{fig:corr-box}
\end{figure}

The first level of comparison assessed whether the estimated respiratory signals followed the reference over time. For each volunteer and method, the z-score-normalized respiratory waveform was paired sample-wise with the belt-derived reference waveform on the common 20~Hz grid, and the Spearman rank correlation coefficient was computed, yielding one value per pair (volunteer, method). The per-subject distributions are shown in Fig.~\ref{fig:corr-box}, and summary statistics are reported in Table~\ref{tab:corr-stats}. 

\begin{table}[!t]
\caption{Per-subject Spearman correlation of the estimated respiratory waveform with the reference belt waveform}
\label{tab:corr-stats}
\centering
\footnotesize
\setlength{\tabcolsep}{3pt}
\begin{tabularx}{\linewidth}{@{}Yccc@{}}
\toprule
\textbf{Method} & \textbf{Median} & \textbf{IQR} & \textbf{Mean $\pm$ SD} \\
\midrule
ToF          & 0.634 & 0.135 & $0.644 \pm 0.100$ \\
Farneb\"ack  & 0.630 & 0.150 & $0.595 \pm 0.137$ \\
Horn-Schunck & 0.634 & 0.156 & $0.587 \pm 0.146$ \\
Lucas-Kanade & 0.644 & 0.148 & $0.601 \pm 0.137$ \\
\bottomrule
\end{tabularx}
\end{table}

All four methods exhibited comparable waveform-level agreement with the reference signal, as shown in Table~\ref{tab:corr-stats}. The median Spearman correlations were closely clustered, with substantial overlap of the interquartile ranges between the methods (Fig.~\ref{fig:corr-box}). The ToF pipeline showed slightly more consistent performance across subjects, reflected in a lower dispersion of correlation values. In contrast, the three RGB-based optical-flow methods each exhibited a single low-correlation outlier corresponding to the same volunteer, who displayed limited thoracic motion during the experiment.

\begin{figure*}[!t]
\centering
\includegraphics[width=0.85\linewidth]{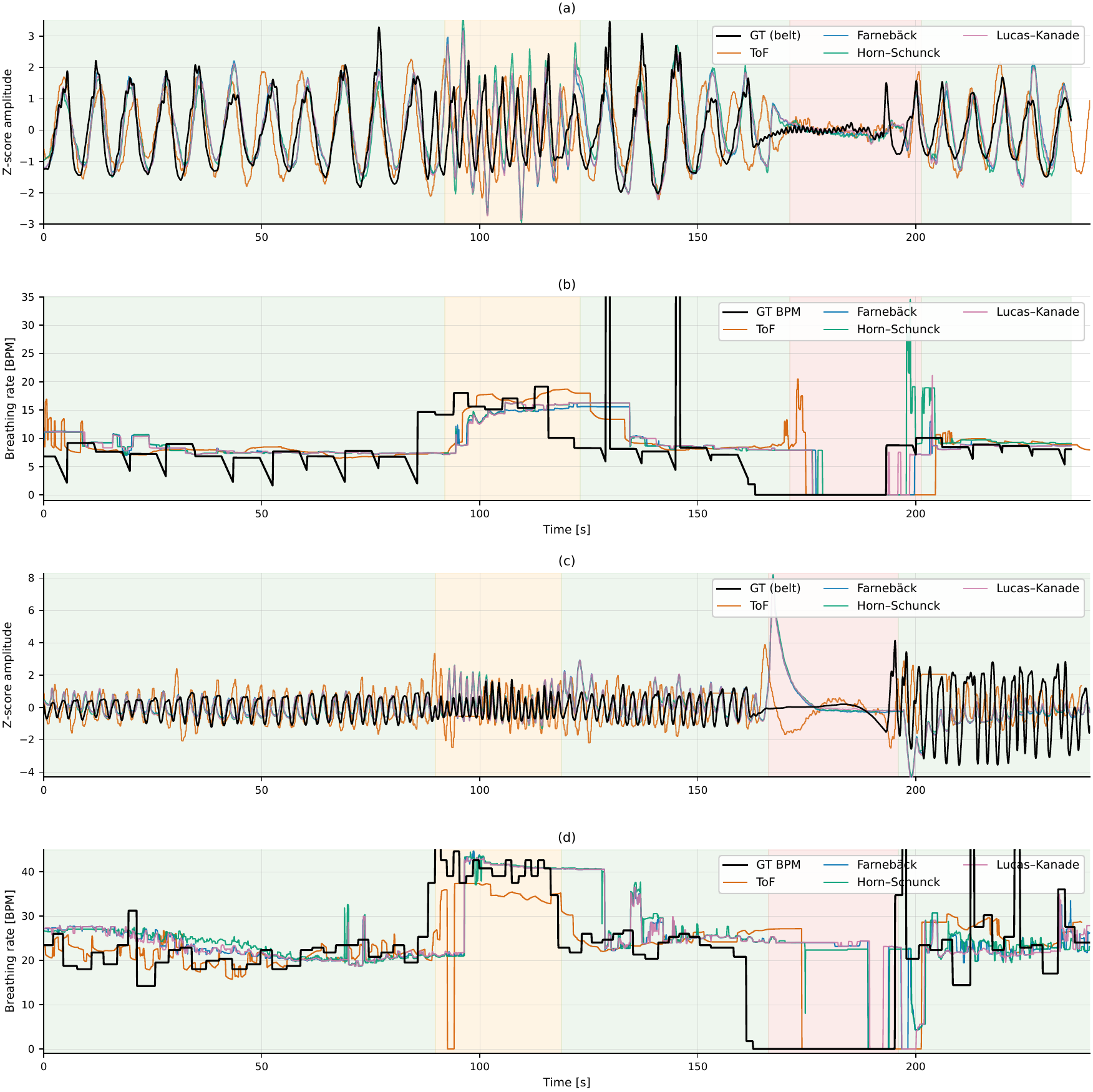}
\caption{Estimated and reference signals for a typical and a challenging recording. \textbf{(a)} Z-score-normalized respiratory waveforms (typical case). \textbf{(b)} Instantaneous RR estimates (typical case). \textbf{(c)} Z-score-normalized respiratory waveforms (challenging case). \textbf{(d)} Instantaneous RR estimates (challenging case). Shaded backgrounds indicate protocol phases: normal breathing (green), hyperventilation (yellow), and breath-hold (red).}
\label{fig:waveforms}
\end{figure*}

To complement population-level statistics, Fig.~\ref{fig:waveforms} illustrates a representative recording in which all methods closely follow the reference signal, along with a more challenging example in which deviations emerge during demanding protocol phases. The two recordings were selected according to a predefined criterion based on the Lucas--Kanade waveform correlation reported in Table~\ref{tab:corr-stats}: the typical case corresponds to the subject whose correlation is closest to the cohort median, and the challenging case to the subject with the lowest correlation, that is, the volunteer with limited thoracic motion mentioned above. These examples are provided for illustration only; the cohort-level evidence is given by the per-subject distributions in Figs.~\ref{fig:corr-box} and~\ref{fig:mae-box} and by the median statistics in Tables~\ref{tab:corr-stats}--\ref{tab:phase-mae}, which are computed over all twelve subjects and are insensitive to individual outlying segments. For visualization, all signals were normalized to a zero mean and unit variance.

\subsection{Respiratory Rate Estimation Evaluation}
\label{sec:res-mae}

\begin{figure}[!t]
\centering
\includegraphics[width=\linewidth]{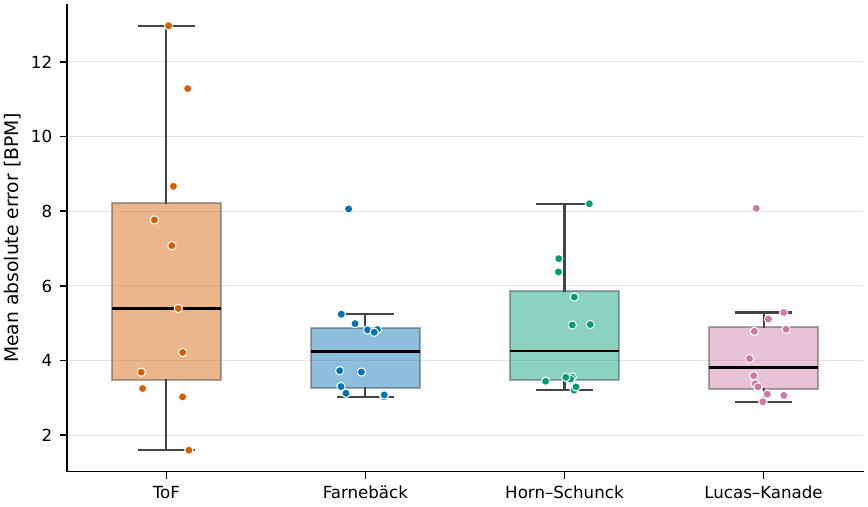}
\caption{Per-subject mean absolute error of the estimated instantaneous RR with respect to the reference belt rate, by method.}
\label{fig:mae-box}
\end{figure}

A second analysis evaluated the absolute accuracy of the RR estimates relative to the reference rate. For each subject and method, the MAE and the signed error were computed throughout the recording. Across the cohort, Lucas--Kanade achieved the lowest median MAE ($3.83$~BPM), followed by Farneb\"ack ($4.24$~BPM) and Horn--Schunck ($4.26$~BPM), while the ToF pipeline yielded $6.24$~BPM. The distributions are shown in Fig.~\ref{fig:mae-box}, with detailed statistics reported in Table~\ref{tab:err-stats}.

\begin{figure*}[!t]
\centering
\includegraphics[width=0.58\linewidth]{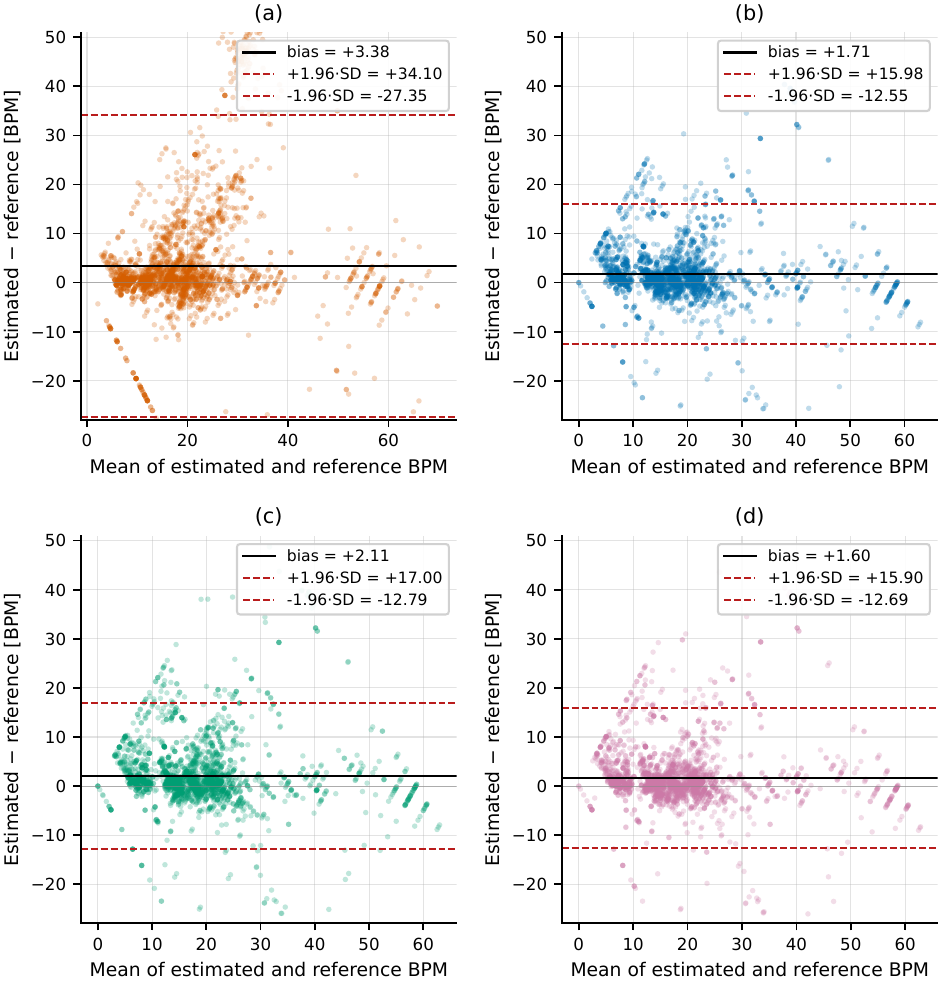}
\caption{Bland--Altman plots comparing instantaneous RR estimates with the reference signal. \textbf{(a)} ToF. \textbf{(b)} Farneb\"ack. \textbf{(c)} Horn-Schunck. \textbf{(d)} Lucas-Kanade.}
\label{fig:bland-altman}
\end{figure*}

A complementary view is provided by the Bland--Altman analysis \cite{MartinBland1986} in Fig.~\ref{fig:bland-altman}. The breath-holding phase was excluded because the reference rate is zero in this regime, and apnea performance is evaluated separately.

The numerical bias values are reported in Table~\ref{tab:err-stats}.
\begin{table}[!t]
\caption{Per-Method RR Error and Bland--Altman Statistics}
\label{tab:err-stats}
\centering
\footnotesize
\setlength{\tabcolsep}{2.5pt}
\renewcommand{\arraystretch}{1.05}
\begin{tabularx}{\linewidth}{@{}Yccc@{}}
\toprule
\textbf{Method} & \textbf{MAE median} & \textbf{MAE mean $\pm$ SD} & \textbf{Bias} \\
                & \textbf{[BPM]}      & \textbf{[BPM]}             & \textbf{[BPM]} \\
\midrule
ToF          & 6.24 & $8.80 \pm 9.41$ & $+3.38$ \\
Farneb\"ack  & 4.24 & $4.39 \pm 1.43$ & $+1.71$ \\
Horn-Schunck & 4.26 & $4.79 \pm 1.65$ & $+2.11$ \\
Lucas-Kanade & 3.83 & $4.29 \pm 1.47$ & $+1.60$ \\
\bottomrule
\end{tabularx}
\end{table}

The three RGB-based methods exhibit a small positive bias and similar limits of agreement, indicating a consistent tendency to slightly overestimate RR. In contrast, the ToF pipeline shows a larger bias and wider limits of agreement, consistent with the heavier upper tail observed in the MAE distribution (Fig.~\ref{fig:mae-box}). This pattern is characteristic of occasional large deviations rather than of a uniform loss of accuracy: the median MAE of the ToF pipeline remains of the same order as that of the RGB methods, while its mean is accompanied by a standard deviation exceeding it, and its limits of agreement are wide despite a comparatively modest bias. The same asymmetry between median and mean is observed within each breathing phase (Table~\ref{tab:phase-mae}), and it coexists with the least dispersed waveform-level agreement of all methods (Table~\ref{tab:corr-stats}), indicating that the extracted depth waveform follows the reference throughout the recording. The most likely origin of these deviations is a brief perturbation of the tracked ROI, during which the depth-derived rate momentarily loses its correspondence with the thoracic motion; this interpretation is supported by the lowest apnea specificity among all methods (Table~\ref{tab:apnea}), which reflects isolated intervals in which the depth-based estimate collapses during active breathing.


\subsection{Breathing Phases Analysis}
\label{sec:res-phase}

\begin{figure}[!t]
\centering
\includegraphics[width=\linewidth]{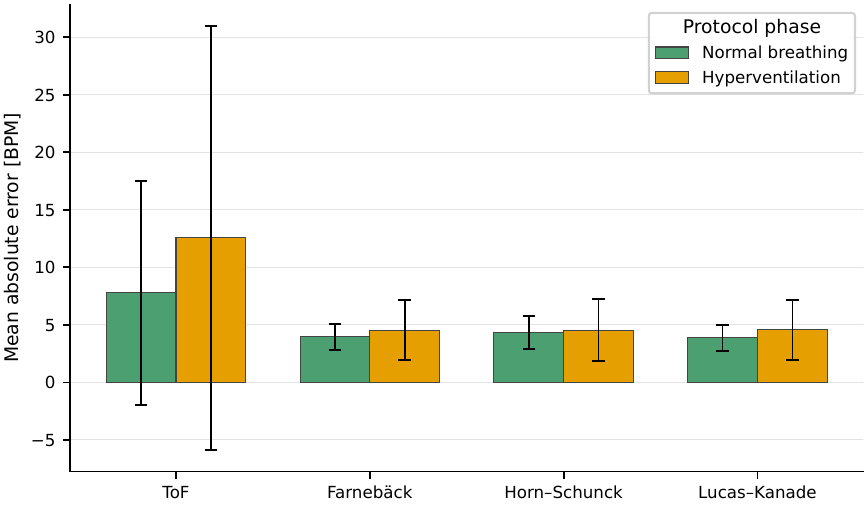}
\caption{Per-subject MAE of the estimated instantaneous RR, split by protocol phase (normal vs. hyperventilation).}
\label{fig:phase-mae}
\end{figure}

The protocol includes three phases: normal breathing, voluntary hyperventilation, and breath-hold. To evaluate performance across different respiratory dynamics, the MAE was recomputed per window for the normal and hyperventilation phases, excluding apnea due to the zero-reference condition. The results are shown in Fig.~\ref{fig:phase-mae}, with numerical values reported in Table~\ref{tab:phase-mae}.

\begin{table}[!t]
\caption{Per-Subject MAE of the Estimated Instantaneous RR, Split by Protocol Phase (Normal vs. Hyperventilation)}
\label{tab:phase-mae}
\centering
\footnotesize
\setlength{\tabcolsep}{2.5pt}
\renewcommand{\arraystretch}{1.05}
\begin{tabularx}{\linewidth}{@{}lYcc@{}}
\toprule
\textbf{Method} & \textbf{Phase} & \textbf{Mean $\pm$ SD} & \textbf{Median} \\
 & & \textbf{[BPM]} & \textbf{[BPM]} \\
\midrule
ToF & normal & $7.78 \pm 9.75$ & 4.23 \\
 & hyperventilation & $12.55 \pm 18.45$ & 3.70 \\
Farneb\"ack & normal & $3.96 \pm 1.15$ & 3.62 \\
 & hyperventilation & $4.54 \pm 2.60$ & 3.94 \\
Horn-Schunck & normal & $4.33 \pm 1.40$ & 3.83 \\
 & hyperventilation & $4.54 \pm 2.68$ & 3.85 \\
Lucas-Kanade & normal & $3.86 \pm 1.11$ & 3.45 \\
 & hyperventilation & $4.56 \pm 2.62$ & 3.90 \\
\bottomrule
\end{tabularx}
\end{table}

The three RGB pipelines exhibit relatively stable behavior between phases, with only a modest increase in MAE during hyperventilation and a limited growth in inter-subject variability, indicating that the optical-flow methods scale reasonably well from resting to elevated RRs. The ToF pipeline shows a markedly stronger phase dependence, with both the mean MAE and the inter-subject variability increasing substantially during hyperventilation. This degradation is driven by the larger thoracic excursions associated with rapid breathing, which exacerbate the sensitivity of the depth-based estimator to small displacements of the tracked ROI.

\subsection{Apnea Detection}
\label{sec:res-apnea}

The breath-hold phase was analyzed as a binary apnea-detection task since the reference RR drops to zero. Each 1~Hz sample was labeled as apnea or non-apnea according to the reference and compared with the pipeline output. Sensitivity, specificity, and $F_{1}$ score are reported in Table~\ref{tab:apnea}.
\begin{table}[!t]
\caption{Apnea-Detection Performance per Method}
\label{tab:apnea}
\centering
\footnotesize
\setlength{\tabcolsep}{2.5pt}
\begin{tabularx}{\linewidth}{@{}Yccc@{}}
\toprule
\textbf{Method} & \textbf{Sensitivity} & \textbf{Specificity} & $\mathbf{F}_{1}$ \\
\midrule
ToF          & 0.660 & 0.951 & 0.662 \\
Farneb\"ack  & 0.530 & 0.983 & 0.647 \\
Horn-Schunck & 0.512 & 0.983 & 0.632 \\
Lucas-Kanade & 0.549 & 0.984 & 0.667 \\
\bottomrule
\end{tabularx}
\end{table}

All pipelines achieve high specificity (0.95--0.98), indicating a low false-alarm rate during active breathing. Differences emerge in sensitivity: the ToF pipeline reaches 0.66, while RGB methods range between 0.51 and 0.55. This reflects the direct measurement of thoracic displacement in depth data, whereas RGB-based estimates depend on the magnitude of motion, which becomes less distinguishable from noise during apnea.

\section{Discussion}
\label{sec:discussion}

The experimental evaluation reported in Section~\ref{sec:results} addresses the central question motivating this work: how RGB optical flow and ToF pipelines compare when implemented on the same low-cost edge platform, applied to the same cohort, and validated against a common reference. The three RGB methods exhibited closely aligned accuracy metrics, whereas the depth-based pipeline showed a higher median error and greater inter-subject variability. In contrast, apnea detection revealed the opposite trend, with the ToF modality achieving a higher sensitivity.

A closer examination of the RGB results clarifies this behavior. Within this modality, the three optical-flow algorithms behaved as practically equivalent estimators of RR. Their interquartile ranges largely overlapped in Figs.~\ref{fig:corr-box} and~\ref{fig:mae-box}, and the limits of agreement (Table~\ref{tab:err-stats}) were also similar, spanning approximately $\pm 13$ to $\pm 17$~BPM around small positive biases of $+1.60$ to $+2.11$~BPM. These findings suggest that the choice of optical-flow algorithm can be guided by secondary criteria such as computational cost or implementation simplicity, without a clear loss in accuracy. Under this criterion, the sparse Lucas--Kanade estimator becomes an attractive option compared to the dense Farneb\"ack and Horn--Schunck methods, since it evaluates motion over a limited set of feature points. This result also addresses one of the gaps identified in Section~\ref{sec:introduction}, namely the lack of side-by-side comparisons under a common protocol.

The apparent discrepancy between waveform agreement and RR accuracy provides further insight into the characteristics of the evaluated modalities. Although the ToF pipeline achieved waveform correlations comparable to, or slightly higher than, those of the RGB methods (Table~\ref{tab:corr-stats}), it also exhibited the largest RR errors (Table~\ref{tab:err-stats}). This finding indicates that preserving the temporal structure of the respiratory signal does not necessarily translate into accurate rate estimation. The depth-based approach consistently captured the overall respiratory pattern, suggesting robust tracking of thoracic motion; however, small deviations in peak localization and rate calculation accumulated into a larger systematic bias ($+3.38$~BPM) and wider limits of agreement. In contrast, the RGB pipelines showed slightly lower waveform agreement but more accurate RR estimates. These observations emphasize that waveform fidelity and rate accuracy assess complementary aspects of performance and should therefore be interpreted jointly. A similar consideration applies to apnea detection, where the preservation of respiratory periodicity and amplitude is more relevant than the precision of individual RR estimates.

In contrast, the depth-based pipeline exhibits more variable behavior. Although its median error remained of the same order of magnitude, both the limits of agreement and the inter-subject variability were larger. Phase analysis helps to explain this difference. Although the median error per subject remained comparable in all breathing conditions, the cohort mean increased markedly during hyperventilation. This effect was driven by recordings with larger thoracic excursions, where increased ROI displacement introduced transient fluctuations in the depth signal. At the same time, the depth-based pipeline achieved the highest sensitivity for apnea detection. This is consistent with the underlying measurement principle: during apnea, the distance signal becomes nearly constant, facilitating detection, whereas RGB methods rely on motion estimates that approach noise levels.

These observations are consistent with the literature discussed in Section~\ref{sec:related-work}. Vision-based studies conducted under highly controlled conditions, such as those of Massaroni et al.~\cite{Massaroni2018} and Romano et al.~\cite{romano2021}, represent a lower bound in terms of error but rely on constrained acquisition setups. The results reported here are closer to those obtained under less controlled conditions, such as the overnight recordings by Zhu et al.~\cite{zhu2019}. For depth sensing, the observed bias aligns with previous ToF- and LiDAR-based studies, although the increased variability likely reflects the inclusion of more demanding breathing regimes, such as hyperventilation.

From an implementation perspective, the real-time measurements in Table~\ref{tab:perf} confirm that both pipelines operate within the constraints of a low-cost embedded platform. The ToF pipeline retains a significantly lower computational demand, whereas the RGB pipeline is primarily limited by pose estimation and dense optical-flow computation. This suggests that both modalities could be executed jointly on the same device, supporting the multimodal approach proposed in this work and addressing the deployment-related gap identified in Section~\ref{sec:introduction}.

Finally, several limitations should be acknowledged. The evaluation was conducted on a relatively small cohort of healthy adults, and RGB and ToF data were acquired during separate recording sessions, precluding direct signal-level fusion and synchronized multimodal analysis. In addition, all measurements were performed under ambient lighting conditions, and apnea episodes were voluntarily induced rather than clinically observed. The study was also not powered to assess sex-related differences. Although the cohort included both male and female participants (Table~\ref{tab:cohort}), the sample size within each subgroup was insufficient for meaningful statistical comparison. Nevertheless, sex-specific anatomical and respiratory characteristics, including differences in thoracic morphology and breathing patterns, may influence the magnitude and distribution of respiratory motion captured by both sensing modalities. These limitations motivate future studies involving larger and more diverse populations, synchronized multimodal acquisition, evaluation under more challenging conditions such as low illumination and partial occlusion, and adequately powered analyzes to investigate potential sex-specific effects on respiratory monitoring performance.

\section{Conclusion}
\label{sec:conclusion}
This work presents a low-cost, real-time platform for non-contact respiratory rate monitoring based on RGB and time-of-flight depth sensing, implemented using a single consumer-grade multimodal sensor on a Raspberry Pi~5 edge device. Three RGB optical-flow methods (Lucas--Kanade, Farneb\"ack, and Horn--Schunck) and a depth-based pipeline were evaluated on a cohort of twelve healthy volunteers under a structured protocol comprising normal breathing, voluntary hyperventilation, and apnea, against a common piezoresistive respiratory belt reference. The evaluation covered waveform agreement, RR estimation accuracy, phase-dependent behavior, apnea detection, and real-time performance.

The RGB optical-flow methods provided consistent waveform tracking and lower RR estimation errors, with Lucas--Kanade achieving the lowest median MAE ($3.83$~BPM) and similar error distributions across the three methods. The ToF pipeline exhibited a higher estimation error (median MAE $6.24$ ~BPM) and greater variability between subjects, but a higher sensitivity in apnea detection ($0.66$ compared with $0.51$-$0.55$ for the RGB methods). Both pipelines satisfied real-time constraints on the target edge device. These results demonstrate that RGB optical flow and depth sensing fulfill distinct and complementary roles within a unified low-cost platform, and that signal-level fusion of the two streams is a natural next step. Future work will focus on validating this approach in broader clinical populations and under challenging conditions such as low-light environments.
\\

\noindent\textbf{Authors' contributions} \textbf{Dominik Jasiówka:} Conceptualization, Methodology, Software, Investigation, Formal analysis, Data curation, Visualization, Validation, Writing -- original draft. \textbf{Ludwin Molina Arias:} Conceptualization, Methodology, Formal analysis, Validation, Supervision, Writing -- review \& editing. \textbf{Patryk Ciężarek:} Investigation, Resources, Validation. \textbf{Joanna Żyźnawska:} Investigation, Resources, Validation. All authors read and approved the final manuscript. 

\vspace{1em} 

\noindent\textbf{Funding} This work was supported by a statutory research subvention from the Polish Ministry of Science and Higher Education (Contract No.~16.16.120.773) and by the National Science Centre, Poland (NCN), under Grant No.~DEC-2025/09/X/ST7/00737. 

\vspace{1em} 

\noindent\textbf{Availability of data and materials} The dataset generated and analyzed during the current study is publicly available in the RODBUK repository of AGH University of Krakow at \url{https://doi.org/10.58032/AGH/EGVPXQ}. The source code for data acquisition, signal processing, and analysis is available on GitHub at \url{https://github.com/ludwin-molina/Contacless-RespiratoryAnalysis}.

\vspace{1em} \noindent\textbf{Declaration of generative AI and AI-assisted technologies in the writing process} During the preparation of this manuscript, generative AI tools were used exclusively for language editing, including grammar correction and improvement of readability. The authors reviewed and approved all revisions and take full responsibility for the final content of the manuscript. 

\section*{Compliance with ethical standards} 

\noindent\textbf{Conflict of interest} The authors declare that they have no known competing financial interests or personal relationships that could have appeared to influence the work reported in this paper. 

\vspace{1em} \noindent\textbf{Ethics approval} All procedures involving human participants were approved by the Research Ethics Committee of the Jagiellonian University Medical College (Resolution No.~47/2026, 22~January 2026) and were conducted in accordance with the principles of the Declaration of Helsinki. 

\vspace{1em} 

\noindent\textbf{Consent to participate} Written informed consent was obtained from all participants prior to their inclusion in the study. 

\vspace{1em} 

\noindent\textbf{Ethics in publishing} This manuscript is original, has not been published previously, and is not under consideration for publication elsewhere.

\section*{Bibliography}
\bibliography{Bibliography}

\onecolumn
\newpage





\end{document}